# Implementation Technologies of an Advanced Cloud-based System for Distribution Operations

**Sotiris P. Gayialis, Evripidis P. Kechagias, Angeliki Deligianni, Grigorios D. Konstantakopoulos, Georgios A. Papadopoulos**

Sector of Industrial Management and Operational Research, School of Mechanical Engineering
National Technical University of Athens
Zografos, 15780, Greece
sotga@mail.ntua.gr, eurikechagias@mail.ntua.gr, aggdelig@mail.ntua.gr, gkonpoulos@mail.ntua.gr, gpapado@mail.ntua.gr

**Abstract**

Today's era is characterized as the "digital transformation era". Digital processes and information systems are used in every aspect of social and business activity. The use of information technology over the internet is so extensive that we interact with it daily without even recognizing it. The technological advances can offer a plethora of improvements for the supply chain processes, especially in the field of distribution planning and execution. At the same time, the development of advanced information systems for usage in urban freight transportation operations is still at an early stage. The scope of this paper is to present the technological content of an advanced routing and scheduling system for transportation and delivery of goods. The system focuses on the routing and scheduling problem in urban areas, as city logistics have become a complex environment for companies to deliver their goods. The presented system deals with both static and dynamic routing and scheduling problems. More specifically, the system can create initial routing plans based on orders, available vehicles, time windows, and traffic forecasting data. Afterward, during the execution of the plans, the system can monitor the fleet, detect deviations from the original plans, and finally, perform rerouting operations when needed. After a brief presentation of the system's modules and functionality, the paper describes thoroughly the technologies used to develop the system. The technological elements of the system are integrated into a cloud environment offering a system that is easy to maintain and can effectively support logistics companies' distribution activities. The system is provided as a Software as a Service with data being maintained on a central host and processed on the cloud. Therefore, logistics companies that decide to implement it can achieve faster, more accurate and more cost-efficient distribution activities while ensuring better customer service.

**Keywords**
Routing and Scheduling, Distribution, City logistics, Software as a Service, Information System

## 1. Introduction

The distribution of goods is a key component of all city supply chains (Sanjay et al. 2020). Urban Freight transportation (UFT) is the part of urban transportation that deals with the distribution of goods in cities and, over the years, has gained significant economic importance (Rose & Martínez 2007). However, as valuable as it is for cities, UFT is associated with a plethora of challenges since many parties are involved. For this reason, UFT should always be considered as a complex set of activities, functions and parties (shippers, receivers, logistics service providers and freight carriers) that all interact with each other and need to be effectively managed (Rose and Martínez 2007; Sánchez-Díaz et al. 2017). UFT is usually carried out on the road due as freight vehicles need to travel through complex urban networks and have direct access to pickup and delivery points. Commonly, UFT is associated with various challenges such as area restrictions (Kin et al. 2017) and traffic congestion (Macharis and Melo 2011) and is responsible for significantly damaging the urban environment (Kijewska et al. 2016; Russo and Comi 2012).

It becomes clear, therefore, that despite the necessity of UFT, it has led to a major transport problem due to the way it has been executed. For this reason, there is a need for innovative ideas concerning the implementation of software technologies in order to achieve more effective, cleaner, and energy-efficient UFT. Technological advances can offer a plethora of improvements for the supply chain processes and especially those related to distribution planning and deliveries execution. It should be noted that, until nowadays, the information systems for routing and scheduling were locally installed on the servers of logistics companies and their main disadvantage was that they didn't have the ability to manage real-time data and perform big data analytics. Therefore, dynamic routing of vehicles was impossible and it was economically unprofitable to supply logistics companies with dynamically changing order and traffic data (Dixit and Chhabra 2015).

However, technological advances have offered the ability to revolutionize the distribution of goods in urban areas. UFT can be aided by utilizing information systems and other appropriate technologies, designed and combined altogether in such a way that they can effectively route and schedule the deliveries of logistics companies. In order to implement such a system, various technologies, such as development platform frameworks, cloud computing services, databases, communication protocols, web services and application programming interfaces (APIs) are needed to be combined and effectively integrated into a single software solution. In this scope, taking into account the challenges of UFT and the need for real-time data management, our team developed, as part of a research project, a cloud-based system for effective and efficient urban freight distributions. The objective of this paper is to present the technologies used for developing the aforementioned system after firstly briefly demonstrating its functionality.

It should be noted that, over the last few years, the way of delivering information systems has shifted from the use of on-premise hardware and local servers to cloud-based solutions. One of the most popular new software distribution models is the Software as a Service (SaaS) model in which the software is licensed on a subscription basis while being hosted centrally and customers gain access through the Cloud. This way of offering software solutions vastly differs from the on-premise software distribution model that was commonly used in the past (Boillat and Legner 2013; Kepes and Subramanian 2020). Via the SaaS distribution model, the software is made available to the customer through the cloud and can be accessed from any device and in any place as long as there is an active internet connection. The charges of this model are only related to licensing fees and can be customized according to the needs of customers (Churakova and Mikhramova 2020; Mel and Timothy 2020). SaaS has gained increased popularity as a software distribution model, as it eliminates the need for high initial setup costs as well as continuous maintenance and upgrade costs. This has contributed to its frequent use in Small and Medium Enterprises (SMEs), while also Large Enterprises (LE) have started considering it as a viable alternative to the existing solutions they use. After all, in both cases, the on-premise needed computational power, is extremely reduced (Churakova and Mikhramova 2020).

This is why, after evaluating alternatives, it was decided that the cloud system for the distribution of goods in urban areas should be offered to end-users as a Software as a Service. However, such an implementation also requires the use of specific technologies for the development of the software. These technologies include the development platform framework, the cloud computing services, the web services, the application programming interfaces, the communication protocols and the databases. In Section 2 of this paper, the functionality and the modules of the advanced routing and scheduling system for urban freight transportation are briefly presented. In Section 3, the integrated technological solutions and their use in the developed system are thoroughly analyzed. Finally, in Section 4 the conclusions, as well as the next steps of the research, are discussed.

## 2. System's Modules and Functionality

### 2.1 System's Architectural Design and Sub-Systems
The created software solution is based on the concept and methodological approach of Gayialis et al. (2018). This approach included a combination of the functions of different subsystems/modules and their interaction in order to create a fully operable system. As already mentioned, the system can be provided by the software company to logistics companies/customers using the Software as a Service (SaaS) distribution model. Since the system is offered as a SaaS, it can be also considered as a distributed system. The basic principle that characterizes distributed systems is their service-oriented architecture (SOA), through which they are divided into smaller subsystems that perform individual functions, with all of them eventually getting grouped together to form an interoperable service (Niknejad et al. 2019).

When gaining access to the system's services, the logistics company can send details of the orders to be planned (delivery points, delivery times, time windows) and the available fleet data (drivers, vehicles, capacity). The system can then combine the above orders and fleet data with historical traffic data and by statistically estimating road traffic, eventually, create an initial deliveries plan. This plan returns to the logistics company and ends up in the navigation devices within the company's vehicles. After this phase is concluded, the distribution activities of the logistics company can begin. However, during the execution of deliveries, the logistics company may need to modify the route of vehicles. These modifications may happen due to cancellations of orders, changes in time windows, mechanical damages to the vehicles as well as unexpected out-of-corporate events such as extreme traffic congestion or weather conditions that may occur during the execution of deliveries. In such cases, if necessary, the system implements rerouting tasks and updates the navigation devices within the company's vehicles with the route modifications based on the changed delivery plans. Therefore, as seen in Figure 1, the architecture of the system consists of two basic subsystems, the routing and scheduling subsystem and the data processing subsystem. Each subsystem includes a set of modules as presented in Figure 1 and described in the next section (section 2.2) of the paper. More details about the algorithms used in the system can be found in Kechagias et al. (2019) and Konstantakopoulos et al. (2020). Description of the methodological approach and the development phase of the system can be also found within the work of Gayialis et al. (2018) and Kechagias et al. (2020a; 2020b).

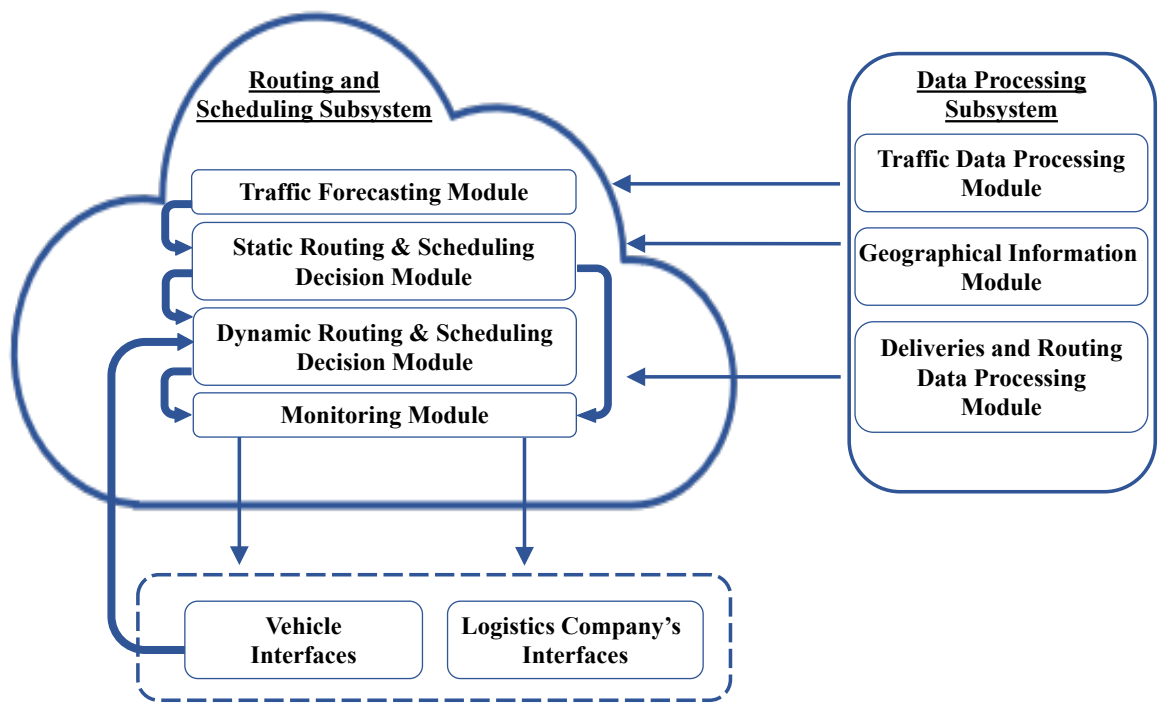

Figure 1: System's Modules and Integration

## 2.1 The Routing and Scheduling Subsystem

The routing and scheduling subsystem includes four modules and allows interactions between the navigation devices on vehicles and the data processing subsystem. The modules of the subsystem are the:
- Traffic Forecasting Module
- Static Routing and Scheduling Decision Module
- Dynamic Routing and Scheduling Decision Module
- Monitoring Module

The traffic forecasting is performed by the traffic forecasting module of the routing and scheduling subsystem. The purpose of this module is to collect historical traffic data and generate a traffic forecast in order to optimize the initial routing plan of the systems' algorithm.

All routing and scheduling data (orders, vehicles and depots) are sent to the decision module for static routing and scheduling. Initially, a statistical estimate of the traffic is performed, and then the initial deliveries plan is generated. The creation of this routing plan is the result of the use of an algorithm that solves the Vehicle Routing Problem with Time Window Limitations (VRPTW). The algorithm's objective function aims to minimize the total deliveries cost according to various constraints. After the routing is concluded, routing plans appear both on the interfaces of the vehicles' navigation devices, and on the logistics company's main interface.

In the routing and scheduling subsystem, there is also the 2nd decision module for dynamic routing and scheduling. This module continuously detects the new information that is received. Depending on the arriving data and by using an algorithm that solves the Dynamic Vehicle Routing Problem with Time Windows (DVRPTW), the module implements either re-routing of orders or modifications of delivery schedules. Eventually, again all changes appear on the navigation devices on vehicles and on the logistics company's main interface.

Finally, the monitoring module of the routing and scheduling subsystem offers the visualization of the routes produced by the vehicle routing modules (static and dynamic). Through this module, the routes are sent either to vehicle devices or to the company's information systems for further analysis . This module can also be used for real-time tracking of vehicle routes. Finally, through this module, the various events, extraordinary or not, that take place during transportation operations and are related to the execution of the itineraries and the deliveries, are recorded.

## 2.2 The Data Processing Subsystem

The data processing subsystem is responsible for data management and processing in order to calculate the static and dynamic plans for the deliveries. The three modules of the subsystem are the:
- Geographical Information Module
- Deliveries and Routing Data Processing Module
- Traffic Data Processing Module

The geographical information module consists of spatial data analysis tools. These tools are able to present the real world at various thematic levels and can also exhibit patterns and relationships between geographic data in order to explore problems, in our case mainly road problems, but are not limited to them. The geographical information module is designed to receive, analyze, interpret, and store data from Global Positioning Systems (GPS) and is integrated with commercial cartographic software for calculations and mapping.

The deliveries and routing data processing module manages the system's database. Indicative data included in this module are static deliveries and routing data, such as deliveries data (customers' name, order quantity, shipping address, cost, dates, payment methods etc.), fleet data (quantity, availability and personal details of drivers and the quantity and capacity of either its private vehicles or the rented etc.), scheduling restrictions (limitations, created by social, technical, environmental, legal and economic factors), distribution times forecasts (historical data from previous distributions and future predetermined factors) and time windows (defined time slots during which distribution is allowed).  However, since the scheduling of deliveries suffers from uncertainty, it is also necessary to gather dynamically changing data, such as order modifications, time window modifications and fleet problems.

The traffic data processing module includes the management of both static and dynamic traffic congestion data. Static traffic data is provided by map providers while traffic forecasting can be performed either by map providers or by the system's algorithms at the traffic data forecasting module. In this way, it is possible to forecast road traffic as well as transportation times from one point to another. In addition, the data provider also offers dynamic (real-time) traffic data coming from all mobile devices in the road network that have active location services. Based on the dynamic data, the module can accurately calculate the speed at which the vehicles move in the road network.

## 3. Implementation Technologies of the Advanced System

### 3.1 Summary of Implementation Technologies

As already mentioned, the system is provided to logistics companies/customers through the Software as a Service distribution model. The system was developed as a Platform as a Service (PaaS), using a development platform framework offered by a cloud computing service. Web services and APIs were used for the management of data exchanged between the central host of the software provider and the cloud platform as well as between the information systems of the users (logistics companies) and the cloud platform, using a communication protocol. Successful communication of the host, the subsystems, and the web services required the use of such an appropriate communication protocol. Finally, a database was also needed for the management and storage of the various data.

The technologies and the technical elements of the system are shown in Figure 2. Additionally, Table 1 presents the specific implemented solutions and the functions of the system that they enable, which are further explained in the next section of the paper.

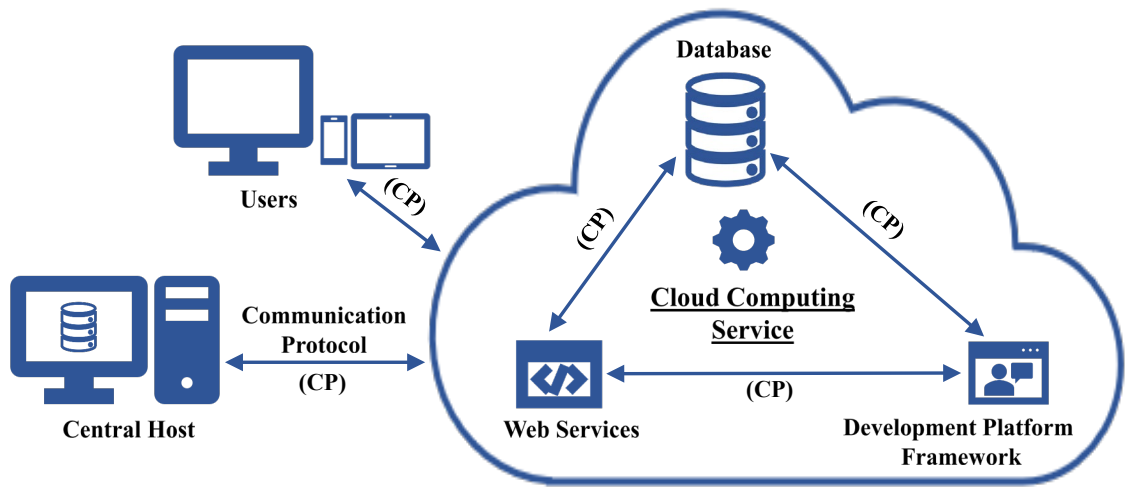

Figure 2: Technologies and Technical Elements of the System

Table 1: Implemented Technological Solutions & Usage in the System's Functions

| Technical Aspect of the System | Solution Implemented | Usage in System's Functions |
|---|---|---|
| Development Platform Framework | Microsoft .NET | Guidance for writing and integrating algorithms and web services into the system |
| Web Services & APIs | Used the ones offered by the geographic and traffic data provider | Receiving all needed services from the geographic and traffic data provider and sending back data for geographic calculations. Receiving user's data for deliveries and sending back the delivery plans. |
| Communication Protocol | REpresentational State Transfer Protocol (REST) | Communication of software host, users, subsystems, web services and APIs |
| Database | Oracle Database | Storage and management of all needed data |
| Cloud Computing Service | Microsoft Azure | Developing the system and providing all necessary tools |

### 3.2 Development Platform Framework

The development platform framework is a library of various integrated technologies and serves as a basis for developing software solutions. The development platform that is created based on the selected framework is essentially the environment in which a piece of software is executed and it can have the form of hardware, operational systems or even web browsers. The development platform contains a back-end and a front-end. The back-end is used to launch the operating system's programs in response to front-end's requests. The implementation of the development platform depends on the needs and the objectives of the software development company (Miller 2003; Vawter and Roman 2020). The development platform provides all tools that are necessary for the operation of the cloud system. The system is developed as a Platform as a Service, allowing developers to create their own applications on an existing platform base with the aid of a plethora of advanced tools.

The two most common platform frameworks for developing applications and web services are Microsoft .Net (commonly found as .Net) and the Java Platform Enterprise Edition (J2EE). Both of these frameworks were born from existing application server technologies (Cobus et al. 2020; Vawter and Roman 2020). For the development of the system, it was decided to use the .NET framework. This framework shows many similarities with the Microsoft Windows operating system and is widely used for business application development. An important advantage of this framework is that developers can more easily create software solutions due to its linguistic neutrality and its simple programming model. Additionally, all vehicle routing and traffic forecasting algorithms, used in the system, were coded in Python, a programming language that can easily "talk" to.NET through its bridges. Last but not least, .NET is also integrated into the cloud computing platform, Microsoft Azure that was used as a service for the development of the system and is analyzed in Section 3.6.

The user interfaces for both the logistics companies and the drivers, that consist the front end of the system, were also created based on the .NET framework. The logistics companies' user interface includes the home page and two additional pages, the deliveries and the assets tabs. In the first tab, users can upload the orders needed to be routed, view the orders and the delivery point on map, and perform the vehicle routing and deliveries scheduling tasks. In the second tab, the users can view and modify data related to their assets. More specifically, they can manage the locations of the hubs, the types of vehicles, the specific vehicles that are available for deliveries, the drivers and finally the shifts of the drivers.

### 3.3 Web Services and APIs

Web services are a subset of web applications, that can be used as independent applications and are published on the Internet. Software development companies can use them as parts of the applications they develop (Curbera et al. 2001; Gottschalk et al. 2018). They can be hosted by any application offering open access and can be used by any client who understands JSON (JavaScript Object Notation) or XML (Extensible Markup Language). Finally, they support the HTTP protocol (Hypertext Transfer Protocol) for URL usage, request/response headers, caching, versions, content formats. Web services can be considered as a subset of application programming interfaces (APIs) that require connection to the internet. APIs are interfaces that can be utilized in order to create software solutions that interact with already developed applications. It should be also noted that web services require the use of a communication protocol, while APIs are protocol agnostic (Souza et al. 2004).

The web services and APIs were used for the static vehicle routing and scheduling of deliveries to:
- receive the geographic data provider's data
- calculate the distance and transit times matrices
- show the created delivery schedules and delivery points on the map
- send the created delivery schedules to the drivers' user interfaces
- receive deliveries, depots and vehicles data from the user
- send the created delivery schedules and plans to the logistic company's system
- send the created delivery schedule to the drivers' portable devices

The web services and APIs were also used for the dynamic vehicle routing and scheduling of deliveries to:

- receive the geographic data provider's data
- receive the traffic data provider's data for real-time traffic forecasting
- recalculate the distance and transit times matrices
- show the created delivery schedules and delivery points on map
- receive delivery exceptions or problems during the execution of the route
- send the recalculated routes and schedules to the logistic company's system
- send the recalculated delivery schedule to the drivers' portable devices

### 3.4 Communication Protocol

In order to communicate and exchange data between the host, the modules of the system, or even in cases where communication with web services and APIs is needed, a communication protocol is necessary (Lo Iacono et al. 2019). The communications are enabled either by using the Simple Object Access Protocol (SOAP) or the REpresentational State Transfer protocol (REST) (Lo Iacono et al. 2019; Potti 2011). For the system's development, it was decided to use the REST protocol as it is the most popular protocol used for web applications due to its flexibility and performance. Additionally, this protocol offers fast response times, a fact that is critical for all real-time functions of the system. Finally, the REST protocol was selected over the SOAP one as it offers live interactions with maps APIs.

### 3.5 Database

A database is a collection of data or organized information (Berg et al. 2012). There are two main categories of databases used in software development, Structured Query Language (SQL) databases and Not only SQL (NoSQL) databases. These two types of databased can be mainly distinguished by the existence or not of relationships between the databases' tables (Berg et al. 2012). The most common SQL databases are MS Access, SQLite, MySQL, SQL Server and Oracle Database (Bassil 2012), and the most common NoSQL database is MongoDB (Berg et al. 2012).

For the development of the system, it was decided to use SQL databases as the system needs to continuously manage the relationships between all stored data. Selecting the appropriate SQL database included the analysis of various commercially available products. Microsoft Access Services was one of those alternatives but wasn't powerful enough and was found to be discontinued by Microsoft on 2018. SQLite is a very light product (less than 500Kb) compared to other solutions but is also unable to withstand the level of data management the system requires. As for the free open-source solution MySQL, it was found that it cannot provide the security and scalability that the system needs. The database that was finally selected for the system was Oracle Database as it can manage dynamic data and also offers great performance and reliability.

### 3.6 Cloud Computing Service

As seen in Section 2 of this paper, the routing and scheduling subsystem is the core element of the developed system and operates in the cloud. For its development, a state-of-the-art cloud computing service offered by the providers of cloud computing was needed. The system could be developed either as a Platform as a Service (PaaS) or as an Infrastructure as a Service. PaaS offers the ability to develop new software solutions on an existing software base or development environment. IaaS, on the other hand, enables the use of standard cloud computing services. For the development of the system, PaaS was found to be the solution that better serves our needs. The software solution selected for the development of the system was Microsoft Azure, which provides all the necessary tools for the integration of algorithms, web services and APIs. Moreover, Microsoft Azure was selected as it can easily communicate with REST APIs and offers .NET as an integrated framework. Finally, Microsoft Azure cloud computing service facilitates the distribution of the system as a Software as a Service (SaaS).

## 4. Conclusions

As it becomes clear, today more than ever, the use of cloud-based software solutions can offer a plethora of benefits for logistics companies and most notably can facilitate their distribution services. In this scope, the presented cloud-based system could aid logistics companies in order to achieve more effective, cost-efficient and environmentally friendly order deliveries to their customers. More specifically, the convenience, reliability and security provided by the system, can lead logistics companies to better control their processes, increase their efficiency while also ensuring the quality of their services/products. All these benefits are vital for logistics companies operating in today's competitive environment. The technologies used for the development and the interconnection of the system and its individual subsystems are the key part for the successful provision of the software solution. The technologies needed for the development of the system included the development platform framework, the cloud computing service, the web services, the application programming interfaces, the communication protocols and the database.

The implementation of these technologies enabled the development of an advanced cloud-based information system for vehicle routing and scheduling, especially in urban areas. The main advantage of the developed system and the implemented technologies is that it can be easily applied in every logistic company, as a cloud service, while it deals with critical constraints of the routing and scheduling problem, like traffic congestions and exceptions during the execution of the schedules. The developed system has been already tested and its operation has been simulated using data from real-life logistics companies. The next goal of our research is to validate the system with the delivery operations of specific logistics companies in real-life conditions. The results of the various performed tests and system validation will be then evaluated and any necessary improvements will be implemented to the system, in order to be ready for exploitation by the logistics sector.

## Acknowledgements

The present work is co-funded by the European Union and Greek national funds through the Operational Program "Competitiveness, Entrepreneurship and Innovation" (EPAnEK), under the call "RESEARCH-CREATE-INNOVATE" (project code: T2EDK-00508 and Acronym: COUNTERBLOCK).

## Biographies

**Sotiris P. Gayialis** has a Diploma in Mechanical Engineering (1997) and a PhD (2008) in Business Process Management and Supply Chain Management from the National Technical University of Athens (NTUA), Greece. He is currently a Teaching and Research Associate of Supply Chain Management Processes in the Sector of Industrial Management & Operations Research at NTUA. He is also an adjunct faculty member of Hellenic Open University. His academic interests are Logistics and Supply Chain Management, Business Process Management, Business Process Improvement, Operations Management and Management Information Systems. He has a twenty year experience in research projects and management consulting. He has participated in a large number of business process improvement and reengineering projects, enabled by IT technology. He has published more than eighty papers in journals, chapters for books and international conferences.

**Evripidis P. Kechagias** is currently a PhD student at the National Technical University (NTUA) of Athens, School of Mechanical Engineering, Sector of Industrial Management and Operational Research. He has studied Mechanical and Industrial Engineering at NTUA (2017) and presented a diploma thesis entitled "Business process management integrated with risk management in the construction industry" that was awarded the honors degree. He also owns a MSc in the field of Technoeconomic studies. His academic interests revolve around the areas of Information Technologies, Business Process Modeling, Analysis and Management, Operational Research, Information System Management, Knowledge Management, Project Management, Industrial Management, Enterprise Resource Planning Systems, Logistics and Supply Chain Management, Production Decisions and Planning. He has published academic and conference papers on these areas.

**Angeliki Deligianni** is currently a PhD student at the National Technical University of Athens (NTUA), in the School of Mechanical Engineering. She holds a Diploma in Mechanical Engineering, with expertise in Industrial Management and Operational Research, and presented a diploma thesis entitled "Software Development Technologies for the Implementation of Urban Freight Transport Systems". Her academic interests revolve around the areas of Risk Management, Management Information Systems, Project Management and Supply Chain Management.

**Grigorios D. Konstantakopoulos** is currently a PhD student at the National Technical University of Athens (NTUA), in School of Mechanical Engineering. He has studied Mechanical Engineering at NTUA (2017), with specialization in Industrial Management and Operational Research, and presented a diploma thesis entitled "Vehicle Routing Systems for Urban Freight Transportation: Field Review and Specifications". His academic interests revolve around the areas of Vehicle Routing Problem, Operational Research, Business Process Management, Logistics and Supply Chain Management. He has published academic and conference papers on these areas.

**Georgios A. Papadopoulos** is a teaching Staff at the School of Mechanical Engineering NTUA Sector of Industrial Management & Operations Research and deals with educational, research and administrative tasks. He has studied Mechanical and Industrial Engineering at NTUA (1994) and followed post-graduate studies at the university of wales, Cardiff business school, UK, obtaining a Master of Business Administration (MBA) (1996). He also obtained a PhD at the National Technical University of Athens, School of Mechanical Engineering, Sector of Industrial Management and Operational Research in the area of Production Planning Decisions (2009). He has participated in many projects and publications in the areas of Production Decisions, Production Planning and Control Systems, Information Systems, Web Design and Development, Application Development.